# Exploring the Smart City Adoption Process: Evidence from the Belgian urban context

Emanuele Gabriel Margherita[a], Giovanni Esposito[b], Stefania Denise Escobar[c], Nathalie Crutzen[b]

[a] *University of Tuscia, Department of Economics Engineering Society and Organization – DEIM, Via del Paradiso, 47, 01100, Viterbo, VT, Italy*
[b] *University of Liège, HEC Liège Management School - Smart City Institute, Rue Saint-Gilles, 35B 4000 Liège, Belgium*
[c] *University of Bozen, Faculty of Economics and Management, piazza Università, 1, 39100, Bolzano, BZ, Italy*

**Abstract**
In this position paper, we explore the adoption of a Smart City with a socio-technical perspective. A Smart city is a transformational technological process leading to profound modifications of existing urban regimes and infrastructure components. In this study, we consider a Smart City as a socio-technical system where the interplay between technologies and users ensures the sustainable development of smart city initiatives that improve the quality of life and solve important socio-economic problems. The adoption of a Smart City required a participative approach where users are involved during the adoption process to joint optimise both systems. Thus, we contribute to socio-technical research showing how a participative approach based on press relationships to facilitate information exchange between municipal actors and citizens worked as a success factor for the smart city adoption. We also discuss the limitations of this approach.

## 1. Introduction

Implementing Smart City programs is a complex technological transformational process leading to profound changes in existing urban regimes and infrastructure components which contributes to increasing quality of life of citizens and reducing $CO_2$ emission in the city [1]. It can be seen as a complex socio-technical system composed of several technologies – information systems, open data access, internet of things – and end-users – citizens [2]. Adopting Smart City is a challenging path for municipal administrations. Previous studies employed a deterministic approach privileging technology functionality over the application domain. However, this approach is not appropriated to meet expected innovation outcomes because users are barely aware of and able to use Smart City technologies. Technological determinism overlooks the role of human agency and prompts a reductionist view of urban technological innovation, which is understood as a mere technical process, taking place outside social, political, organisational, and cultural settings [3]. Therefore, recent studies reject such a reductionist view of urban technological innovation and recommend to develop Smart City projects as socio-technical projects where the interplay between technologies and users are simultaneously considered [2]. It requires a joint optimisation of both social and technical systems through a participative approach [4] enabling municipalities to interact with the end-users of the Smart City technologies – citizens - throughout the adoption process. As explained by Mumford [4], the effective adoption of technology - within a given domain - is possible when users are involved in the adoption process. Here, users can understand the technology and/or contribute to its design [5].

Our study advances knowledge regarding participative approaches and its success factors for Smart City development based on a single case study of a Belgian City that is implementing a Smart City program.

We address the following research question: *"How does a municipality employ a participative approach for the adoption of a Smart City?"*

We contribute to socio-technical research presenting a socio-technical approach over a domain where often existing studies use a techno-determinist perspective. We illustrate the actions that municipalities put into practice to reach the joint optimisation of both systems highlighting the limits of this participative approach. We also propose future research avenues to develop Smart City research from a socio-technical perspective.

## 2. Smart City

Smart city initiatives are complex transformational processes [6] leading to profound modifications of existing urban regimes and infrastructure components [1]. Such initiatives have gained momentum as a suitable approach to rule, manage, govern and lead the city. These can be understood as urban strategies that seek to advance technological solutions to the pressing sustainability issues facing policymakers [7]. Particularly, their aim is to integrate ICTs into urban systems in order to improve citizens' participation in the city governance and to enhance their quality of life through more efficient, user-oriented and sustainable transport and energy networks [8]. Becoming smart is indeed an ambition which an increasing number of cities are trying to achieve. Examples of strategies for supporting smart city development can be found all over the world and smart city researchers have made significant efforts to investigate their design and implementation process [9].

The effective adoption of Smart City technologies requires a socio-technical approach acknowledging the interrelatedness of social and technical aspects of sustainable urban transition processes [10]. From this perspective, the interaction of social and technical factors creates the conditions for successful sustainable urban transition processes that occur when socio and technical elements are jointly optimised, that is, technologies effectively work and users effectively employ these technologies [11].

A participative approach is an effective way to reach the conjoint optimisation of both systems [4, 12]. It refers to an approach which involves the users of the adopted technologies along the adoption phase. This approach invites city governments to refuse the deterministic, one-size-fits-all approach to the development of sustainable urban technologies [13]. Smart City literature shows three participative approach: the double, triple or quadruple helix models of collaboration. In the double-helix model, the interaction is only between ICT enterprises - selling their smart technologies to public authorities - and local governments buying smart technologies from the private sector [14]. This model generates an entrepreneurial mode of governance in which information technology corporations working in the market of smart city services become the main providers of ICT solutions to solve urban problems. According to [15 pg. 67], these public–private collaborations are supposed to allow "businesses to pursue their own interest whilst […] serving collective interests and public value". However, research suggests that this collaborative model does not provide the collective intelligence which is necessary to drive smart city development and to face the complexity that this socio-technological transformation process poses [16]. One solution can be the triple helix model referring to a collaboration model in which interactions occur between university, industry and government, to foster smart city solutions. Alternatively, a much more open and inclusive collaborative ecosystem can rely on the quadruple-helix model in which all the city stakeholders representing governments, universities and businesses are involved, along with citizens and civil society organisations. Such quadruple model has the potential to lead to successful smart city developments through co-creation practices, participatory governance, community-led urban development, open innovation, crowdsourcing and user-driven innovation and [17].

## 3. Research Method

A single case study is a fruitful methodology to explore the novel phenomenon, yet not well studied [18, 19]. This methodology consists of triangulating different data sources. We make use of semi-structured interviews. We carried out 24 semi-structured interviews following a purposive criterion sampling approach [19]. Interviews were conducted between May 2018 and June 2018, whereas three follow-up interviews were conducted in 2020. We interviewed key informants of the Smart City adoption [20]. Our criterion consisted in interviewing the main actors of the adoption process of the Smart City technologies: (1) public authorities, (2) firms, and (3) research and education institution and (4) civil society associations. The interviews explored the following topics: actors' reasons for their involvement in Aplhaville's smart city projects, their roles in the projects, their a-posteriori evaluations and expectations, and next steps in project development.

Interview data were triangulated with several documents covering the 2012-2019 period, which covers the design of the Smart City and the adoption process of the Smart City [21]. These documents are retrieved from reliable sources, which are the strategic plan of the city, urban and regional policy documents, and press articles from national newspapers. We analysed the data following the procedure for the qualitative enquiry by Corbin and Strauss, making use of first level and second level coding [22].

## 4. The case of a Belgian Smart City

The Belgian city - that we call Alphaville to preserve the anonymity of interviewees - has a population of around 110.000 citizens. The most important economic sector in the city is the service industry - especially tourism - and (contrarily to several Belgian cities) production factories are located outside the city. To attract tourists, the city maintained its historical center and small roads to move around the city. Over the last years, the city experienced problems ranging from mobility to issues of socio-economic development:

*"There are mobility problems, (…) as well as problems in terms of cultural creativity and economic development. When I was elected mayor and I mentioned the fact that we had to "dare" the creative city and become a Smart City, I was not understood by everyone. I was always fascinated by urban technologies - I have university diploma in ICTs law - and I wanted to adopt them in Alphaville (…) to foster the sustainable development of our city from different points of view such as energy, employment, mobility, supporting local entrepreneurship through creative hubs* (Mayor)."

*"The Smart City is about urban development, particularly from an economic point of view. The key problem we need to address is 'how can we capitalise on our urban strengths to prepare the city of tomorrow in terms of dynamic entrepreneurship?* (Academic partner)"

The newly elected mayor – with a professional background in management of ICT projects – decided to address those problems through a Smart City program consisting of 3 main projects: (1) an Intelligent Transport Systems (ITS), that will increase safety and will tackle Alphaville's $CO_2$ emission and traffic congestion problems. These systems make use of geolocalisation and sensors [23] equipped on roads and parking to acquire mobility data. Data are stored in the cloud [24], analysed and showed to citizens on panels around the city; (2) a creative hub for the development of local start-ups in the ICTs sector; and, (3) the construction of a smart neighbourhood to host collective intelligence sessions involving local start-ups, citizens and public institutions aimed at planning the urban development of Alphaville. The program was funded with the financial support of the European Union. Throughout the program development, concerned municipal officers communicated the initiative to business stakeholders and citizen association by means of press conferences aimed at sharing their smart city vision with the general public. The general public had doubts about the usefulness of the program as they could not comprehend how these technologies will have positively impacted the city. To secure the support of the general public, the municipal administration established contacts with journalists and set up a mission to visit a French Smart City to which Alphaville would have resembled once accomplished the program. Journalists were invited to be part of this mission so that they could better understand the municipal vision behind the programmed Smart City transformation of Alphaville and explain it to the general public:

*"In my head it's clear how the smart city will contribute to the development of Alphaville. I think this is not the case for the general public and it is normal because these places do not yet exist physically* (Business partner).*"*

*"Few journalists initially understood the Smart City concept. They saw it as incidentally connected to urban development and therefore tended to see our program as superfluous. They couldn't appreciate how this could become the backbone of local policy action (...). We had thus to persuade them and we set up a delegation to visit the Smart City of Betaville in France* (Mayor).*"*

*"Journalist did not understand and their articles were incorrect (...) so we took them with us in the delegation visiting the Smart City of Betaville in France. (...) Betaville was really an example of what we were going to do. (At the end of the visit) everyone got it. Coming out of the Betaville visit, I remember a journalist asking me "why didn't you say us all these things earlier?", but we did. They only understood it by going there and seeing what was happening* (Urban planning officer).*"*

Although some doubts remained among some citizens, such initiative with the press succeeded to secure the support of the general public and to align key actors, with positive impact on the implementation of the smart city program in Alphaville.

*"It took a while to align all key actors. The mayor leadership was pivotal in all this (...). After visiting Betaville, the majority of the opponents that participated in the delegation – including journalists, rectors, professors and businesses - aligned around the same understanding of smart city* (Business partner).*"*

*"(T)he Smart City will improve the quality of life in our city. It makes everyday life easier. So there is no reason to go back* (Association of local businesses).*"*

*"The Smart City is still a buzzword. It is a trendy theme that the mayors and its administration try to use for their own purpose (...) I think that in Alphaville there is still a need to ensure that citizens understand and appropriate the Smart City for their own purposes. This is not obvious. It is perceived as a hype concept in the interest of Start-Up, etc. I am not sure that all the population is really aware of what is happening* (Association of local citizens with interest in digital issues).*"*

At the time we are writing, the implementation of the three Smart City initiatives is effectively underway. The creative hub has come to an end and begins to carry out its activities with the local ICT entrepreneurial community. The ITS is in the phase of testing with citizens involved in the testing process through a mobile app which allow them to find the fastest itinerary to reach and provide real-time information of traffic. The construction of a smart neighbourhood to host collective intelligence sessions is ongoing and citizens are regularly informed about the ICT solutions that will be deployed to collect and synthesise policy inouts to be transferred to the local administration.

## 5. Findings & Discussion

The case shows that the effective adoption of the Smart City technologies passes through a participative approach based on press relationships between the municipalities and the general public. This approach allowed the municipal administration to secure the support of citizens (i.e. end-users) in the implementation of three smart city projects aimed at solving different socio-economic problems and increasing their quality of life. Press relationships were particularly aimed at improving the understanding of the general public about the ongoing smart city program as the general public had doubts about the usefulness of the program and could not comprehend how the planned smart technologies will have positively impacted the city life. Press relationships aimed at strengthening a shared understanding of technological urban deployment can be thus considered as an important success factor, likely to have a positive impact on the implementation of smart city technologies.

This participative approach has limitations. The municipality opted for an approach that does not inform the citizens directly but uses the intermediation of press officers such journalists. In our case study, the design of smart technologies mainly driven by political leadership (i.e. the mayor) and was supported by business actors through indirect involvement of citizens. Such involvement limits to information sharing with no citizen participation in key decision-making phases. Our case study fits into Drapalova and Wegrich's [25] 'business-politics coalition' model of smart city development where a key role is played by political leaders who are highly interested in applying smart technologies to urban governance and service delivery. Because the smart agenda is politicised and driven by political actors, the business does not directly impose their plan to local administrations but execute or complement the political program.

Building on a public governance approach to smart cities and political economy theories of business–politics relations, Drapalova & Wegrich [25] identify three more varieties of smart city innovation models along two major dimensions, namely the magnitude of civic mobilisation (i.e. direct involvement of citizens) and the degree of politicisation (i.e. direct involvement of elected politicians). Besides the 'business-politics coalition' model that we found in our analysis, they mention: citizen-centred, disjointed and captured city models. Further smart city research should investigate more these three models and, particularly, the citizen-centred one where smart technologies are 'co-created' by governments in collaboration with both firms and civil society actors. Urban technological deployment is indeed a socially-constructed process whereby governments, firms and civil society actors may attribute different meanings to technology. Such meanings may sometimes be contradicting, thus preventing actors from establishing a shared common understanding of future city challenges and possible solutions. When this happens, cooperation among actors may fail with negative consequences on the outcome of the urban innovation process [26]. Therefore, co-creation practices governments, firms and civil society actors are pivotal factors in order to advance technological solutions that simultaneously meet global and local needs while implementing a widely-shared and desired urban development. In this regard, the concept of Public-Private-People partnership (4P) is one emerging innovative institutional arrangement to facilitate the involvement of private actors and the general public ("people") in a joint process of urban decision-making involving issues from legislation as well as local and national. Unlike most traditional statist and market-based arrangements [27–29], 4P institutional arrangements bring the focus on citizens as co-producers of policies and on appropriate institutional frameworks within which "the policies that guide society are the outcome of a complex set of interactions involving multiple groups and multiple interests ultimately combining in fascinating and unpredictable ways" [23 pg. 533]. Such institutions facilitate inter-organisational processes between multiple network actors, including government, (profit and not-for-profit) businesses and citizens.

Additionally, our study indicates that, besides business and political actors, university actors are involved thus suggesting that in our case, the triple helix innovation model also applies. Therefore, a fruitful research avenue is to explore how to involve users in the design and usage of a Smart City. A traditional method, such as the focus group can achieve this purpose. Also, innovative ways like e-participation can facilitate this process. For instance, the creation of a website, where users can write their thoughts or compile a survey to express their accordance or disagreement with the project, can be a valid means. Then, on-line or in-presence training, which is an appropriate method in the organizational context [31], can be proposed to citizens to increase their knowledge related to the use of these technologies.

Finally, other studies can put emphasis on the study of the role of the facilitator introduced by Mumford for the participative approach that is an external consultant, with a neutral status, which aims at designing the new socio-technical system and adopting the new technology through the fulfilment of the needs of the end-user of the technologies and the steering committee of the adoption. How does this role change in the domain of Smart City? How can a facilitator harmonise a socio-technical system where a multitude of actors with distinct needs is involved?

## 6. Conclusion

In this study, we present a socio-technical perspective of Smart City adoption. Although most of the study employs a technical perspective which privileges the technical functionalities over the social norms during the adoption of the Smart City, our study explains how a Belgian municipality uses a participative approach involving users along with the adoption. The case illustrates the adoption process of the Smart City is influenced through a participative approach based on relationships with the press. Echoing Drapalova & Wegrich (2020) [25], our findings suggest that citizens' mobilisation and political leadership are primary factors that can influence urban technological innovation processes. We additional show the involvement of university actors. Nevertheless, we explain that our case study does not allow to understand smart city development in the context of direct citizen participation. Citizens participation is an important topic as their reluctance and/or overt opposition to novel services, or products can be an important barrier to public sector innovation. That is likely to emerge when government organisations are not able - or do not want - to respond to societal challenges and civil society demands [26, 32]. We thus invite researchers to investigate further the topic of participative approaches to Smart City development.